# Facilitating Bioinformatics Reproducibility

Christopher R. Keefe[1], Matthew R. Dillon[1], Chloe Herman[1,3], Mary Jewell[2], Colin V. Wood[1], Evan Bolyen[1], J. Gregory Caporaso[1]

1. Center for Applied Microbiome Science, Pathogen and Microbiome Institute, Northern Arizona University, Flagstaff, AZ, USA.
2. Department of Epidemiology, University of Washington, Seattle, WA, USA
3. School of Informatics, Computing and Cyber Systems, Northern Arizona University, Flagstaff, AZ, USA

## Abstract

Study reproducibility is essential to corroborate, build on, and learn from the results of scientific research but is notoriously challenging in bioinformatics, which often involves large data sets and complex analytic workflows involving many different tools. Additionally many biologists aren't trained in how to effectively record their bioinformatics analysis steps to ensure reproducibility, so critical information is often missing. Software tools used in bioinformatics can automate provenance tracking of the results they generate, removing most barriers to bioinformatics reproducibility. Here we present an implementation of that idea, Provenance Replay, a tool for generating new executable code from results generated with the QIIME 2 bioinformatics platform, and discuss considerations for bioinformatics developers who wish to implement similar functionality in their software.

## Introduction

Reproducibility, the ability of a researcher to duplicate the results of a study, is a necessary condition for scientific research to be considered informative and credible.[1] Peer review relies on study documentation to maintain the trustworthiness of scientific research.[2–4] Without comprehensive documentation, reviewers may be unable to verify a study's validity and merit, and other researchers will be unable to interrogate the results or learn from the researchers' approach, limiting the study's value.

The biomedical research community has recently been concerned with a "reproducibility crisis," and several high-profile publications have shown researchers unable to confirm findings of original studies.[5,6] This discussion generally focuses on one type of reproducibility failure: an inability to corroborate a study's results. However, this literature neglects a deeper issue: many studies fail to provide even the minimum necessary documentation to reproduce a study's methodology.

Although there is no standard nomenclature for reproducibility in the literature, existing classification systems illustrate the goals of different types of reproducibility. For example, the Turing Way defines research as "Reproducible," "Replicable," "Robust," or "Generalizable" based on whether a study's results can be repeated using methods and data that are the same as, or different from, the original study.[7] Gundersen and Kjensmo create similar categories in their work on reproducibility, but they define a hierarchy based on the degree of generality.[8]

By imposing the Gundersen and Kjensmo hierarchy on the Turing Way categories (Figure 1), we contextualize key factors in generalizability within the broader goals of reproducible research. While it can be important to corroborate results using the same methods and data, findings that are reliable and generalizable to other settings are more useful and important than findings that are valid only in their own specific context.[9] Generalizable studies ultimately provide the building blocks for more impactful scientific conclusions.

High quality research documentation allows researchers to both corroborate original work and expand towards more generalizable conclusions, making documentation essential to the goal of reproducibility. Analytic software tools that also produce research documentation have the potential to reduce the risk of paper retraction, facilitate collaboration, review, and debugging, and improve the continuity and impact of scientific research.[7] Suggestions of criteria that these tools should aim for is presented in Keefe 2022.[10]

In bioinformatics, research typically involves large datasets, complex computer software, and analytical procedures with many distinct steps. In order to reproduce such a study, one needs both prospective provenance, the analytic workflow specified as a recipe for data creation, and retrospective provenance, the details of the runtime environment and the resources used for analysis.[11] Prospective provenance typically involves a document written by the user, for example using Jupyter Notebooks or RMarkdown, which includes annotation and executable code. Retrospective provenance can be achieved by capturing information about the analysis as it is run, including hardware and software environments, resource use, and the data and metadata involved. Most published research does not meet these needs, as researchers must balance competing demands on their time and grapple with publication structures that incentivize producing new work over documenting for reproducibility.[7,9,12]

QIIME 2 is a biological data science platform that was initially built to facilitate microbiome research[13], but has been expanding into domains including analysis of highly-multiplexed serology assays[14] and pathogen genomics.[15] QIIME 2 has a built-in system that tracks data provenance for users as they run their analyses, and the popularity of this feature is in part responsible for its adoption in other domains. In QIIME 2, users conduct analysis using Actions that each produce one or more Results. The prospective and retrospective provenance of all preceding analysis steps are automatically stored in each Result, allowing users to determine how a Result was generated and an analysis was conducted (Figure 2), even if scripts or notes were not recorded (or were misplaced) by the user.

The work presented here attempts to improve computational methods reproducibility in bioinformatics by reducing the practical overhead of creating reproducibility documentation. Built around the automated, decentralized provenance capture implemented in QIIME 2, we present *Provenance Replay*, a tool that validates the integrity of QIIME 2 Results, parses the provenance data they contain, and programmatically generates executable scripts that allow for the reproduction, study, and extension of the source analysis.

# Methods

Provenance Replay is written in Python 3.8,[16] and depends heavily on the Python standard library, NetworkX,[17] pyyaml,[18] and QIIME 2 itself, from which it takes advantage especially of the PluginManager and the Usage API. It ingests one or more QIIME 2 results and parses their provenance data into a Directed Acyclic Graph (DAG), implemented as a NetworkX DiGraph. It produces outputs by subsetting and manipulating this DiGraph and its contents. Outputs include BibTeX-formatted[19] citations for all Actions and Plugins used in a computational analysis, and executable scripts targeting the user's preferred QIIME 2 interface. Users interact with the software through a command-line interface implemented with Click,[20] or using its Python 3 API.

The initial software design was based on literature review, existing API targets, and discussion with QIIME 2 developers, as well as an initial requirements engineering process and formal focus groups with prospective users. This process is described in detail in Keefe 2022.[10] Provenance Replay is supported by QIIME 2 versions 2021.11 and newer, and can parse data provenance from Results generated with any version of QIIME 2. As of QIIME 2 2023.5, the software is included in the QIIME 2 "core distribution" (such that it is installed with QIIME 2). Prior to the QIIME 2 2023.5 release, Provenance Replay can be accessed in the QIIME 2 development environment (see https://dev.qiime2.org for install instructions).

Provenance Replay is capable of replaying a single QIIME 2 Result in a few seconds, and a very large analysis (450 results) in 8-10 minutes on a contemporary small-business laptop (Intel Core i7-8565U CPU @ 1.8GHz, 16 GB RAM, OpenSUSE Tumbleweed running on an M.2 SSD). As such, most users will not need to work in a cluster environment, and native installation is recommended (and supported on Linux, MacOS, and Windows via Windows Subsystem for Linux 2 (WSL2)).

The provenance-lib software is open source and free for all use (BSD 3-clause license) and is available at https://github.com/qiime2/provenance-lib.

# Results

Provenance Replay is software for the documentation and enactment of *in silico* reproducibility in QIIME 2, which can produce command-line (bash) and Python 3 scripts directly from a QIIME 2 Result. Provenance Replay outputs are self-documenting, using universally unique identifiers (UUIDs) to identify them as products of specific QIIME 2 Results, and they include step-by-step instructions for users to execute the scripts produced. Provenance Replay also implements MD5 checksum-based validation of Result provenance, which can alert if their Results were altered since they were generated, in which case the data provenance would no longer be reliable.

Provenance Replay meets many of the goals outlined in Keefe 2022[10] for reproducibility tools:

- Completeness: Provenance Replay provides comprehensive access to all captured provenance data.
- Ease of Documentation: users can generate a complete "reproducibility supplement," including replay scripts and citation information, with a single command through different user interface types.
- Ease of Reproduction: Replay scripts are executable with minimal modification, target a variety of interfaces, and are self-documenting.
- Readability: Replay documents are designed for humans, formatted nicely, and include their own usage instructions.
- Learning-readiness: A variety of common interface-specific script output formats are provided, reducing interface-related barriers to readability and learning.
- Interpretation and communication: By providing multiple user interfaces for Provenance Replay, as well as multiple target interfaces for its outputs, users with varied computational skills can translate analyses for one another.

Provenance Replay automatically removes most barriers to *in silico* methods reproducibility in QIIME 2 (with some exceptions discussed below). This simplifies the process of documenting research, and it has already been used to generate reproducibility supplements for scientific publications.[21]

# Discussion

Comprehensive study documentation is a necessary prerequisite to scientific reproducibility, but many researchers are unable to provide adequate documentation due to limited training, resources, and competing demands for their time. Provenance Replay largely automates *in silico* reproducibility in QIIME 2, and this approach can provide a model for other scientific computing platforms.

When designing these systems, it is important to keep in mind that positive identification of study data and metadata is essential to being able to corroborate study results, and the level of expertise among consumers of study documentation varies widely. These issues should be considered in the design of provenance tools.

In addition to prospective and retrospective provenance, *reproducible* and *robust* bioinformatics (Figure 1) involves unambiguous identification of the data used in an analysis. This can be challenging to achieve, as it requires stable database identifiers, and if data can be mutated after data submission, the identifiers should be versioned. QIIME 2 uniquely identifies its data artifacts with UUIDs, and those artifacts are immutable (once created, they can not be changed without the creation of a new data artifact with a different UUID). However, data artifacts would also need to be centrally located and accessible to generally allow others to corroborate study results, and that is currently outside the scope of QIIME 2. Ensuring positive identification of study data to facilitate re-analysis should be a target for systems aiming to provide reproducible bioinformatics. QIIME 2 may ultimately address this by enabling programmatic access to data artifacts through Qiita[22], or integrating q2-fondue[23] commands with Provenance Replay results, to load data from the NCBI Sequence Read Archive into QIIME 2 as a step in replaying an analysis.

Systems that aim for improved analysis documentation should facilitate interpretation by consumers with different computational backgrounds. For example, if a goal is to support complete methods descriptions that are generally accessible, it can help for reproducibility documentation to be presented for the different interfaces that are available to the underlying computational tools. For example, QIIME 2 can be accessed through a Python 3 API, a command line interface (CLI), and through various workflow systems, including Galaxy. At present, Provenance Replay can provide Python 3 scripts, Jupyter Notebooks, and bash scripts, providing different documentation options for users of the API and CLI. A future development target for Provenance Replay is to provide reproducibility instructions for Galaxy[24] users as well. Providing complete documentation through higher-level (e.g., graphical) interfaces is more verbose, but ultimately expands the audience who can learn from that documentation.

Tools such as Provenance Replay provide a means for ensuring study reproducibility while reducing the documentation burden on bioinformatics users, who may forget to record steps in their computational lab notebooks, or who may not be aware of all of the information that needs to be documented due to lack of training. Designing computational tools that remove this burden from the user by recording data provenance for them is possible and will be a major step forward in methods reproducibility, allowing researchers to more easily corroborate results, learn from the work of others, and build on the conclusions of scientific studies.

# Figure captions

**Figure 1**: Turing Way reproducibility classes are ordered in a Gundersen-like hierarchy based on generality. Each class represents a situation in which the same results are produced using either the same or different data and the same or different analysis methods.

**Figure 2**: Diagram of the provenance of a QIIME 2 Artifact. Left: a directed, acyclic graph tracing this history from initial data import into QIIME 2, through the creation of a visualization for analytical interpretation or publication, from top to bottom. Right: the "action details" captured during the creation of the two nodes highlighted in blue.

# Author contributions

Christopher R. Keefe: Conceptualization, Investigation, Software, Writing – original draft, Writing – review & editing

Matthew R. Dillon: Conceptualization, Supervision, Writing – review & editing

Chloe Herman: Validation, Writing – review & editing

Mary Jewell: Writing – original draft, Writing – review & editing

Colin V. Wood: Writing – review & editing, Software

Evan Bolyen: Conceptualization, Supervision, Writing – review & editing

J. Gregory Caporaso: Conceptualization, Supervision, Funding acquisition, Writing – review & editing

# List of works cited

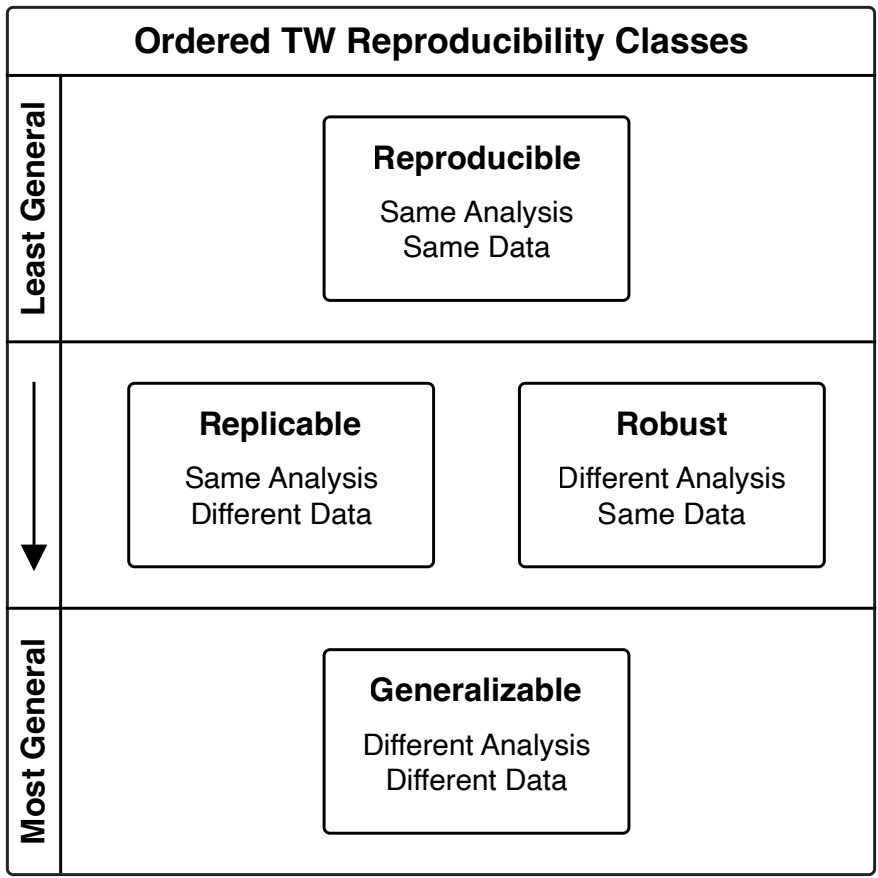

Figure 1.

Figure 2.